\documentclass[aps,pra,twocolumn,showpacs,superscriptaddress,longbibliography]{revtex4-2}

\usepackage{amsmath,amssymb,amsfonts,psfrag}
\usepackage{bm}
\usepackage{color}
\usepackage{soul}
\usepackage[dvips]{graphicx}
\usepackage[unicode]{hyperref}
\hypersetup{
    pdftitle={Scaling Dynamics},
    pdfauthor={Ashton Bradley},
    colorlinks=true,
    linkcolor=blue,
    citecolor=blue,
    filecolor=black,
    urlcolor=blue,
    backref=section
}

\def\qq{d}
\def\urlprefix{}
   \def\url#1{}

\def\del{\partial}

\def\bpp{\mathbf{p}}
\def\bPP{\mathbf{P}}
\def\rr{\mathbf{r}}
\def\bxx{\mathbf{x}}
\def\bqq{\mathbf{q}}

\def\ii{i}
\def\jj{j}

\newcommand{\ali}[1]{\begin{align}#1\end{align}}
\newcommand{\eref}[1]{Eq.~(\ref{#1})}
\newcommand{\eeref}[1]{(\ref{#1})}
\newcommand{\fref}[1]{Fig.~(\ref{#1})}
\newcommand{\sref}[1]{Section~\ref{#1}}

\begin{document}

\title{Scaling dynamics of the ultra-cold Bose gas}
\author{Ashton S. Bradley}
\affiliation{Dodd-Walls Centre for Photonic and Quantum Technologies, Department of Physics, University of Otago, Dunedin, New Zealand}
\author{Jordan Clarke}
\affiliation{Dodd-Walls Centre for Photonic and Quantum Technologies, Department of Physics, University of Otago, Dunedin, New Zealand}
\author{Tyler W. Neely}
\affiliation{Australian Research Council Centre of Excellence for Engineered Quantum Systems, School of Mathematics and Physics, University of Queensland, St. Lucia, QLD 4072, Australia.}
\author{Brian P Anderson}
\affiliation{Wyant College of Optical Sciences, University of Arizona, Tucson, AZ 85721, USA}
\date{\today}

\begin{abstract}
  The large-scale expansion dynamics of quantum gases is a central tool for ultra-cold gas experiments and poses a significant challenge for theory.
  In this work we provide an exact reformulation of the Gross-Pitaevskii equation for the ultra-cold Bose gas in a coordinate frame that adaptively scales with the system size during evolution, enabling simulations of long evolution times during expansion or similar large-scale manipulation. Our approach makes no hydrodynamic approximations, is not restricted to a scaling ansatz, harmonic potentials, or energy eigenstates, and can be generalized readily to non-contact interactions via the appropriate stress tensor of the quantum fluid. As applications, we simulate the expansion of the ideal gas, a cigar-shaped condensate in the Thomas-Fermi regime, and a linear superposition of counter propagating Gaussian wavepackets. We recover known scaling for the ideal gas and Thomas-Fermi regimes, and identify a linear regime of aspect-ratio preserving free expansion; analysis of the scaling dynamics equations shows that an exact, aspect-ratio invariant, free expansion does not exist for nonlinear evolution. Our treatment enables exploration of nonlinear effects in matter-wave dynamics over large scale-changing evolution.

\end{abstract}
\maketitle
\section{Introduction}
The expansion of quantum gases is a central interrogation tool in experiments, being the canonical approach for extracting information about atomic momentum distributions. Absorption imaging after a period of free expansion has provided many insights into the properties of trapped ultra-cold gases, including playing a decisive role in the first observations of quantum degeneracy~\cite{anderson_observation_1995,demarco_onset_1999}. Expansion imaging has been used to probe many phenomena in excited Bose-Einstein condensates (BECs), including quantum vortex dynamics~\cite{neely_observation_2010,anderson_resource_2010,serafini_vortex_2017,bisset_observation_2017-1,dalfovo_optical_2018-1} and quantum turbulence~\cite{henn_emergence_2009,neely_characteristics_2013,navon_emergence_2016}; it has also been used as a quantum simulator of cosmological effects such as inflation dynamics~\cite{eckel_rapidly_2018}, Hawking effects~\cite{fedichev_gibbons-hawking_2003}, and particle production~\cite{fedichev_cosmological_2004}. 

The simplest scenario is one of linear evolution, for which the one-body dynamics map the initial momentum distribution into the far-field position distribution. Interactions complicate this mapping, rendering it non-exact in general, although some special cases are tractable analytically. In Fermi gases expansion interactions can often be neglected~\cite{ketterle_making_2008-1}, or solved exactly~\cite{castin_exact_2004}, while in a BEC, interactions dominate the scaling evolution in the Thomas-Fermi regime~\cite{castin_bose-einstein_1996,kagan_evolution_1996}. Interaction effects can also be somewhat mitigated via a short interval of rapid expansion along a tightly confined axis to reduce the particle density~\cite{weiler_spontaneous_2008}. However, a complete understanding of expansion dynamics remains an important open problem.

The theoretical challenge involves describing the dynamics with sufficient resolution over large changes in system size and long evolution times. 
Existing theoretical approaches favour a hydrodynamic formulation, applied to the 3D collisionless Boltzmann gas~\cite{guery-odelin_mean-field_2002}, the Tonks regime~\cite{minguzzi_exact_2005}, Fermi gases~\cite{menotti_expansion_2002,castin_exact_2004,deng_superadiabatic_2018,deng_observation_2018,deng_observation_2016}, degenerate Bose-Fermi mixtures~\cite{hu_expansion_2003}, and BECs~\cite{modugno_effective_2018,viedma_effective_2020}. The central tool is  the scaling ansatz for the wavefuction~\cite{kagan_evolution_1996,castin_bose-einstein_1996,castin_exact_2004,del_campo_frictionless_2011,campo_shortcuts_2012}, leading to dynamics of the Ermakov type for the scaling parameters~\cite{ermakov_transformation_1880,werner_unitary_2006-1}; this approach has been employed to great effect to construct shortcuts to adiabaticity (STA)~\cite{rohringer_non-equilibrium_2015,schaff_fast_2010,schaff_shortcut_2011,guery-odelin_shortcuts_2019}. Ultimately, these works rely on scale invariance, which applies to a class of systems with scaling dimension of the interaction potential equal to that of the Laplacian~\cite{del_campo_frictionless_2011,gritsev_scaling_2010}; the scale invariant formulation can be used to analyze STA for a class of monomial potentials~\cite{deffner_classical_2014}, including the special case of time-dependent harmonic confinement. A wide range of other external potentials have been treated in STA; for a review see~\cite{guery-odelin_shortcuts_2019}. Expansion dynamics have also been studied using numerically intensive large-scale direct simulation of the Gross-Pitaevskii equation (GPE), giving insights into vortex imaging~\cite{dalfovo_free_2000-1} and rapidly rotating condensates~\cite{simula_observations_2005}.

Here we take an alternative approach to the general problem of scale changing dynamics in the GPE. We combine a scaling-like transformation of the wavefunction with a physical definition of the scaling parameters in terms of system size parameters. In contrast with previous works that solved for scaling parameters approximately or via an ansatz, this enables an exact reformulation of the GPE in coordinates that dynamically adapt to changes in the system size.  The resulting equations of motion for the scale parameters account for internal stresses of the quantum fluid that drive scaling dynamics. Our formulation can describe cold scalar BECs with $s$-wave interactions manipulated by arbitrary external potentials. As our reformulation only relies on the existence of a fluid continuity equation and stress tensor, a broader class of quantum fluids such as spinor and dipolar systems may be handled via the appropriate stress tensor~\cite{chomaz_long-lived_2019,kawaguchi_spinor_2012,tanzi_observation_2019}. 

The paper is structured as follows: in \sref{sec:background}, we briefly outline the GPE theory of the ultra-cold Bose gas. In \sref{sec:eom} we develop the scaling dynamics formulation of the GPE. In \sref{sec:app}, we test our formulation on the ideal gas and Thomas-Fermi regimes, and identify a regime of linear aspect-ratio invariant expansion. In \sref{sec:outlook}, we offer some perspectives on our findings, and point to interesting future directions. 

\section{Background}\label{sec:background}

\subsection{Gross-Pitaevskii equation}
At temperatures well below the BEC transition a system of identical bosonic atoms is well described by the Gross-Piteavskii equation~\cite{pitaevskii_bose-einstein_2003}. Choosing the convenient normalization
\begin{align}
  \int d^\qq r\;|\psi(\rr,t)|^2=1 
\end{align}
for $N$ atoms in $\qq$ spatial dimensions, the GPE Hamiltonian is 
\begin{align}
  H=\int d^\qq r\left(\frac{\hbar^2}{2m}|\nabla\psi|^2+V|\psi|^2+\frac{g_d N}{2}|\psi|^4\right),
\end{align}
where $V(\rr,t)$ is the external trapping potential, and $g_d$ is the $s$-wave interaction parameter reduced to dimension $d$ by integrating over the tightly confined transverse state (assumed separable) in $3-d$ dimensions. Hamilton's equation for $\psi$, 
\begin{align}
  i\hbar\frac{\partial \psi}{\partial t}&=\frac{\delta H}{\delta \psi^*},
\end{align}
gives the GPE
\ali{\label{gpedef}
i\hbar\frac{\del\psi(\rr,t)}{\del t}&=\left(-\frac{\hbar^2\nabla^2}{2m}+V(\rr,t)+g_d N|\psi(\rr,t)|^2\right)\psi(\rr,t).
}
Local particle conservation enforces continuity
\begin{align}\label{cont} 
\partial_t|\psi|^2+\partial^\jj J_\jj=0,
\end{align}
where the current density is $J_\ii\equiv i\hbar/(2m)(\psi\partial_\ii\psi^*-\psi^*\partial_\ii\psi)$, and $\partial_\ii\equiv\partial/\partial x_\ii$. We will make use of tensor index notation and the Einstein summation convention, however there is a trivial metric: there is no difference between raised and lowered indices. 

The superfluid current density evolves according to the equation of motion~\cite{frisch_transition_1992,winiecki_pressure_1999}
\begin{align}\label{jeom}
  \partial_t J_\ii+\frac{1}{m}\partial^\jj T_{\ii \jj}+\frac{1}{m}|\psi|^2\partial_\ii V=0,
\end{align}
with fluid stress tensor
\begin{align}\label{gpest}
  T_{\ii\jj}&\equiv\delta_{\ii\jj}\frac{g_d N|\psi|^4}{2}+\frac{\hbar^2}{4m}\left((\partial_\ii\psi^*)(\partial_\jj\psi)-\psi^*\partial_{\ii\jj}\psi+\textrm{c.c.}\right).
\end{align}
Equations \eeref{jeom}, \eeref{gpest} are exact for the GPE, and similar expressions can be derived for bosonic quantum fluids with different interaction potentials. Noteably for our purposes, the current and stress tensor are free from pathologies that arise at zeros of the wavefunction (e.g. vortex cores) in a hydrodynamic formulation using the Madelung transformation. In what follows we avoid hydrodynamics and use the GPE stress tensor, \eref{gpest}, to develop an exact reformulation of the GPE; we find that the diagonal terms of the stress tensor generate a fluid pressure that plays a central role in BEC expansion dynamics. 

\subsection{Dynamical Coordinate Scaling}\label{sec:dcs}
To derive a GPE suitable for describing expanding systems of ultra-cold Bose atoms we formulate the dynamics in a scaled coordinate system, where in general the scaling can take on any time dependence~\cite{castin_bose-einstein_1996,kagan_evolution_1996}. We introduce scaled coordinates  
\begin{align}\label{xlamdef}
  x_\ii &\equiv\rho_\ii(t)\lambda_\ii(t),
\end{align}
   where both the coordinates $\rho_\ii$ and scaling parameters $\lambda_\ii$ depend upon time, in such a way to render the laboratory frame time-independent. We have, for example
\ali{
  \dot x_\ii\equiv \frac{dx_\ii}{dt}&=\dot\rho_\ii\lambda_\ii+\rho_\ii\dot\lambda_\ii=0,
}
without summation. The scaled coordinates evolve according to
\ali{\label{dtex}
  \dot\rho_\ii&= -\rho_\ii\frac{\dot\lambda_\ii}{\lambda_\ii
},
}
so that a scaling expansion corresponds to a coordinate contraction that preserves the product $\rho_\ii\lambda_\ii$. 

The scaled wavefunction $\phi(\rho_\ii,t)$ conserves probability under rescaling 
\begin{align}
  \int d^\qq x\;|\psi(x_\ii,t)|^2&=\int d^\qq \rho\; |\phi(\rho_\ii,t)|^2=1.
\end{align} 
 We use the component shorthand $\phi(\rho_\ii,t)$ to represent $\phi(\rho_x,\rho_y,\rho_z,t)$, and similarly for $\psi(x_\ii,t)$. The coordinate transformation defines a rescaling of $\phi$ up to a local gauge transformation. We define the transformation to a scaled wavefunction $\phi(\rho_\ii,t)$ as
\ali{\label{psiscale}
\phi(\rho_\ii,t)&\equiv\bar\lambda(t)^{\qq/2}\psi(\rho_\ii\lambda_\ii,t)\exp{\left(-\frac{im}{2\hbar}\rho_\jj\rho^\jj\dot\lambda_\jj\lambda^\jj\right)}.
}
in $\qq$ spatial dimensions, allowing for arbitrary $\lambda_\ii(t)$, with initial conditions $\lambda_\ii(0)\equiv 1$. The geometric mean $\bar\lambda(t)\equiv (\Pi_{\ii=1}^\qq \lambda_\ii(t))^{1/\qq}$ enforces probability conservation. The convenient phase removes the phase gradients generated by scaling~\cite{castin_bose-einstein_1996}. Note that we have not imposed any constraints on the wavefunction: this definition does not enforce a scaling ansatz as the field $\phi(\rho_i,t)$ is unconstrained by the transformation; however the dynamics of $\lambda_i$ and $\phi$ remain to be determined.  
\section{Scaling dynamics: equations of motion}\label{sec:eom}
Thus far we have summarized well known results required for our reformulation of the GPE, to which we now turn. To make use of the definition \eeref{psiscale} we also require a definition of the scaling parameters in terms of $\psi(x_i,t)$. We work in laboratory coordinates initially centered on the mean position of the condensate~\footnote{A shift in the center of mass, comparable to the system size scaling, should also be treated by a dynamical shift in coordinates by $\langle x_i\rangle(t)$. Here we focus on the scaling problem.}. The parameters characterizing the system size are then  
 \begin{align} 
  a^2_\ii(t)&\equiv \int d^\qq x\; x_\ii^2|\psi(x_\ii,t)|^2.
 \end{align}
We define the scaling parameters in terms of the system size parameters as
\begin{align}\label{lamdef}
  \lambda^2_{\ii}(t)&\equiv \frac{a^2_\ii(t)}{a^2_\ii(0)},
\end{align}
defined such that $\lambda_i(0)\equiv 1$.
In the scaling coordinates the quadratic moments 
\begin{align}
  \int d^\qq\rho\; \rho_i^2|\phi(\rho_i,t)|^2&=a_i^2(0),
\end{align}
are time-independent by definition. Hence, whatever form the new GPE for $\phi(\rho_i,t)$ takes, the system size is time invariant in our chosen coordinates defined by \eeref{xlamdef} and \eeref{lamdef}. These definitions form an essential feature of the scaling dynamics formulated below.

Our aim now is to find equations of motion for the scaling parameters $\lambda_\ii(t)$, and wavefunction $\phi(\rho_\ii,t)$, for an arbitrary time-dependent trapping potential $V(x_\ii,t)$. We can do this now that we have an appropriate definition of the scaling parameters in terms of the time-dependent system size. To keep notation light, we use the shorthand $a_\ii\equiv a_\ii(0)$ for initial widths of the BEC, hereafter declaring any time dependence explicitly. It will also be convenient to reduce all dynamical equations to first order in time, introducing additional variables $\sigma_\ii(t)\equiv \dot\lambda_\ii(t)$ for the rates of change of the scaling parameters. 

We start by differentiating \eref{lamdef}, using continuity, and integrating by parts (discarding surface terms), to find
\begin{align}\label{dl0dt}
  \frac{d\lambda_\ii }{dt}=\frac{1}{\lambda_\ii a_\ii^2}\int d^\qq x\;x_\ii J_\ii,
\end{align}
the equation of motion for the scaling parameters in terms of the field $\phi(\rho_\ii,t)$. We do not need to evaluate this directly due to our reduction to first order. Differentiating \eref{dl0dt} and using \eref{jeom} to replace $\partial_t J_i$, we find the equation of motion
\begin{align}\label{ds0dt}
  \frac{d\sigma_\ii}{dt}&=-\frac{\sigma_\ii^2}{\lambda_\ii}+\frac{1}{\lambda_\ii ma_\ii^2}\int d^\qq x\; \left(T_{\ii\ii}-|\psi|^2x_\ii\partial_\ii V\right).
\end{align}
Note that the off-diagonal terms in the stress tensor do not contribute since 
\begin{align}
  \int d^d x\; x_\ii \partial^jT_{ij}=-\delta_{ij}\int d^d x\; T_{ii}.
\end{align} 

We can write the equation of motion as 
\begin{align}\label{dsdt}
  \frac{d\sigma_\ii}{dt}&=\frac{f_\ii+\tau_\ii-\sigma_\ii^2}{\lambda_\ii},
\end{align}
containing the force exerted by the trapping potential  
\begin{align}\label{pforce}
  f_\ii&\equiv-\frac{1}{ma_\ii^2}\int d^\qq x\;|\psi|^2x_\ii\partial_\ii V,
\end{align}
the stress tensor term 
\begin{align}\label{sttrace}
  \tau_\ii&\equiv\frac{1}{ma_\ii^2} \int d^\qq x\; T_{\ii\ii},
\end{align}
and the inertial term $\propto \sigma_i^2$.
The diagonal elements of the fluid stress tensor generate hydrostatic pressure~\cite{davidson_turbulence_2015} in equilibrium, and in this dynamical setting they provide forces driving scaling dynamics. The scaling equation of motion, \eeref{dsdt}, holds for any quantum fluid stress tensor and external potential, providing a general description that can accommodate beyond $s$-wave systems, such as spinor or dipolar interactions, via the appropriate stress tensor of the fluid. 

Hereafter we specialize to the GPE stress tensor, \eref{gpest}. We can simplify further by integrating by parts and discarding surface terms to give the equivalent form 
\begin{align}\label{tausimp}
  \tau_\ii&=\frac{1}{ma_\ii^2}\int d^\qq x\;\left(\frac{g_d N}{2}|\psi|^4 +\frac{\hbar^2}{m}|\partial_\ii\psi|^2\right),
\end{align}
showing kinetic and interaction contributions to the superfluid pressure. 
Equation (\ref{dsdt}) describes the GPE dynamics of the scaling parameters for arbitrary initial conditions and external potentials. Our remaining task is to find an equation of motion for $\phi$.

Differentiating with respect to time, including the explicit coordinate time dependence, we have total time derivative
\ali{
\frac{d\phi}{dt}&=\frac{\partial \phi}{\partial t}-\frac{\dot\lambda_\ii}{\lambda^\ii}\rho^\ii\partial_\ii\phi,
}
where we used \eref{dtex}, and $\del_\ii\phi\equiv \partial\phi/\partial \rho^\ii$ is the derivative in $\rho^\ii$ coordinates. After some algebra, given in Appendix \ref{app:a}, we find $\phi(\rho_\ii,t)$ evolves according to a GPE-like equation similar to that found by Castin and Dum~\cite{castin_bose-einstein_1996}, except that we now have a completely general time-dependent external potential. To complete the transformation to scaling coordinates, we transform \eref{dsdt}, and use \eref{ddmu} to simplify the kinetic term~\footnote{The integral of the final term in \eref{ddmu}, appearing in \eref{sttrace}, exactly cancels the $-\sigma_\ii^2/\lambda_\ii$ term in \eref{ds0dt}.}, and arrive at a coupled system of first-order time evolution equations for our complete set of dynamical variables $\lambda_\ii(t),\sigma_\ii(t),\phi(\rho_\ii,t)$. Writing all equations and the initial conditions in terms of the scaling coordinates, we arrive the scaling dynamics equations (SDE) 
\begin{widetext}
\begin{align}\label{dlamdt}
  \frac{d\lambda_\ii}{dt}&=\sigma_\ii(t),\\
  \frac{d\sigma_\ii}{dt}&=\frac{1}{\lambda_\ii(t) m a_\ii^2}\int d^\qq\rho\;\left(\frac{g_d N}{2}\frac{|\phi|^4}{\bar\lambda(t)^\qq}-|\phi|^2\rho_\ii\partial_\ii V(\rho_\ii\lambda_\ii(t),t)+\frac{\hbar^2}{m\lambda_\ii(t)^2}|\partial_\ii\phi|^2 \right), \label{dsigdt}\\
  i\hbar\frac{\partial \phi}{\partial t}&=\left(-\frac{\hbar^{2} \partial_{\ii}^{2}}{2 m \lambda^\ii(t)^2}+V( \rho_\ii\lambda_\ii(t),t)+\frac{g_d N}{\bar{\lambda}(t)^\qq}|\phi|^{2}+\frac{m}{2}  \dot{\sigma}_{\ii}(t) \lambda^{\ii}(t)\rho_\ii\rho^\ii\right) \phi,\label{dphidt}
\end{align}
\end{widetext}
starting from the initial conditions 
\begin{align}
  \label{ic1}
  \lambda_\ii(0)&=1,\\\label{ic2}
  \sigma_\ii(0)&=\frac{1}{a_\ii^2}\int d^\qq \rho\;\rho_\ii J_\ii(\rho_\ii,0),\\\label{ic3}
  \phi(\rho_\ii,0)&=\psi(\rho_\ii,0)\exp{\left(-i\frac{ m\rho_\jj \rho^\jj}{2\hbar}\sigma_\jj(0) \right)},
\end{align}
in response to the arbitrary external potential $V(x_\ii,t)$.

We have derived a complete reformulation of the GPE in the form of a scaling GPE containing scaled kinetic, potential, and interaction terms. Equations \eeref{dsdt}, and \eeref{dlamdt}---\eeref{ic3} are our main results. To the best of our knowledge such a reformulation, starting from~\eeref{xlamdef}, \eeref{psiscale}, and \eeref{lamdef} has not appeared in the literature. In our derivation we made no hydrodynamic or other approximations. As we show in \sref{sec:app}, the scaling GPE is able to accommodate arbitrary changes in scale by adapting with the system size. Relying only on the quantum fluid stress tensor and the volume current density, our treatment also provides a general starting point for scaling dynamics in a range of quantum fluids with different interparticle interactions. 

To conclude this section, note that in scaling coordinates the GP energy may be found using the transformed Laplacian (\ref{ddmu}), giving
\begin{align}
  H&=\int d^\qq \rho\;\Bigg(\frac{\hbar^2|\partial_\ii\phi|^2}{2m\lambda^\ii(t)^2}+V|\phi|^2+\frac{g_d N}{2}\frac{|\phi|^4}{\bar\lambda(t)^\qq}\notag\\
  &\quad\quad\quad\quad+\frac{m}{2}\sigma_\ii(t)\sigma^\ii(t)\rho_\ii\rho^\ii|\phi|^2\Bigg).
\end{align}
In the case of free expansion,  $H$ is a constant of the motion, and the initial kinetic, interaction and trap energy transforms into energy of expansion stored in the steady expansion rates, $\sigma_\ii(t)\to\bar\sigma_\ii$ reached in the long time limit. This contrasts with the lab frame where $H$ is simply the kinetic energy.

\section{Applications}\label{sec:app}
Our aim is to verify that the scaling GPE reproduces well-known behavior of the GPE, and discuss some interesting special cases that may be easily described by the scaling GPE. 

We initially proceed numerically: we simulate the free expansion of an ideal gas with anisotropic initial confinement, and a highly prolate system in the Thomas-Fermi regime. We consider applying a strong parabolic antitrapping potential to extract the position distribution. Finally, we consider the situation relevant to quantum turbulence experimental measurements~\cite{henn_emergence_2009}, by considering conditions where an anisotropic initial state can evolve while preserving its aspect ratios.

In the following we solve the first order system of equations using the DifferentialEquations.jl~\cite{rackauckas_differentialequations.jl_2017} package written in Julia~\cite{bezanson_julia_2017}, and plot the results using Makie.jl~\cite{danisch_makiejl_2021}.
\subsection{Ideal gas expansion} 
We first consider the free expansion of an ideal gas. The full dynamics can be found analytically, as developed in detail in Appendix \ref{app:b}. Here we use this linear evolution as a test of numerical simulations of the scaling GPE. 

We consider the implications of the scaling GPE dynamics after long expansion times. In particular, the aspect ratios of the trapped system should eventually invert. For a non-interacting system, the dynamics approach a steady state solution for \eref{dsdt}, given by the balance of the stress tensor term with the inertial term: $\sigma_\ii^2=\tau_\ii$. The system reaches a finite steady-state rate of expansion, $\bar\sigma_\ii$, which takes the following form in lab coordinates:
\begin{align}
  \bar\sigma_\ii^2=\frac{\hbar^2}{m^2a_\ii^2}\int d^\qq x \; |\partial_{\ii}\psi(t\to\infty)|^2.
\end{align}
During free expansion of an ideal gas 
\begin{align}
  i\hbar\partial_t\psi&=-\frac{\hbar^2}{2m}\nabla^2\psi,
\end{align}
the kinetic term is only a quadratic phase winding in momentum space. The momentum-space wavefunction $\tilde\psi(p_\ii,t)$, is then simply 
\begin{align}
  \tilde\psi(p_\ii,t)&=\exp{\left(-i\frac{ p_\jj p^\jj t}{2m\hbar }\right)}\tilde\psi(p_\ii,0),
\end{align}
and the steady expansion rate $\bar\sigma_\ii$ is set by the \emph{initial} kinetic energy
\begin{align}
 \bar\sigma_\ii^2&=\frac{1}{m^2a_\ii^2}\int d^\qq p\;p_\ii^2   |\tilde\psi(p_\ii,0)|^2.
\end{align} 
For the harmonic oscillator ground state with oscillator lengths $\alpha_\ii\equiv(\hbar/m\omega_\ii)^{1/2}$, $\langle p_\ii^2\rangle = \hbar^2/(2\alpha_\ii)^2$, and $\langle x_\ii^2\rangle = \alpha_\ii^2/2$, the scaling dynamics approach the constant rate $ \bar\sigma_\ii=\omega_\ii$. 

\begin{figure}[!t]
  \centering
\includegraphics[width=\columnwidth]{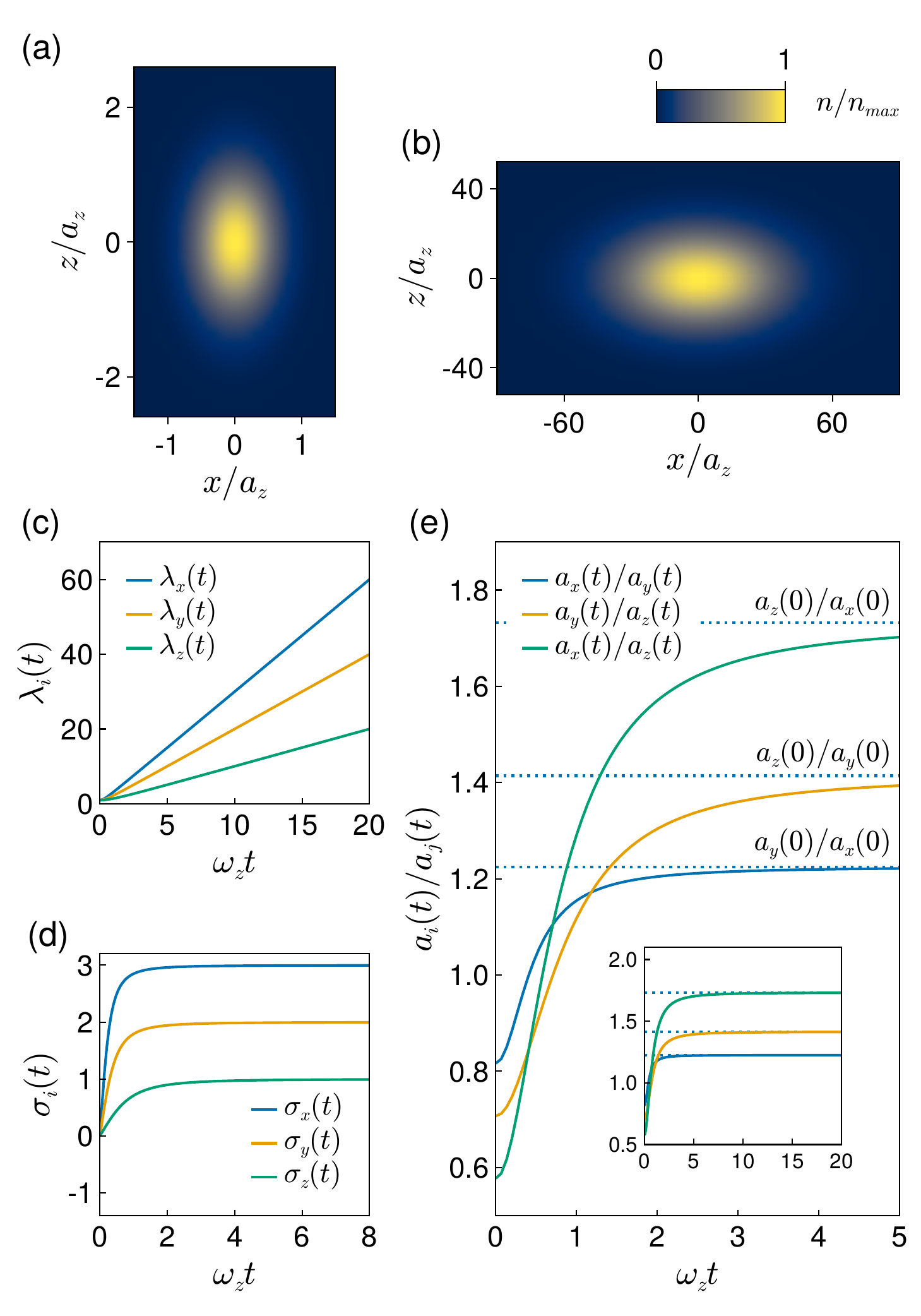}
\caption{Simulated expansion dynamics of the ideal Bose gas using Eqs.~(\ref{dlamdt}-\ref{ic3}) on a grid of $64^3$ points. The ground state in a harmonic trap with $(\omega_x,\omega_y,\omega_z)=(3,2,1)$ has initial density in the $x-z$ plane shown in (a), evolving to (b) after free expansion for $t=20/\omega_z$. (c) scaling parameters rapidly become linear in time; (d) the scaling derivatives approach the initial trap frequencies $\sigma_\ii(t)\to\omega_\ii$; (e)  position aspect ratios invert on timescale $\sim 1/\omega_z$, and approach the initial momentum aspect ratios for $t\gg 1/\omega_z $ (inset). 
}
\label{fig1}
\end{figure}

We numerically simulate ideal gas expansion for a system with trap frequencies $(\omega_x,\omega_y,\omega_z)=(3,2,1)$. We work in length and time units of $\alpha_z$ and $1/\omega_z$ respectively. We evolve an initial anisotropic domain $(L_x,L_y,L_z)=5(\alpha_x,\alpha_y,\alpha_z)$, with $(N_x,N_y,N_z)=(64,64,64)$ points. Integrating the scaling dynamics equations numerically up to $t=20/\omega_z$, we observe the evolution shown in Fig.~\ref{fig1}. The aspect ratios are seen to invert for long expansions, with timescale set by $1/\omega_z$, the longest trap period of our chosen parameters. The aspect ratio inversion is due to the well-known mapping of initial momentum information into the final position distribution. Note that the equations are numerically stable, allowing simulation of arbitrary expansion times. 

As a final and essential check on the scaling GPE formulation, in Appendix \ref{app:b} we show analytically that the momentum distribution is mapped into the position distribution in the long-time limit of linear evolution.  

\subsection{Expansion of a Thomas-Fermi ground state} 
The scaling coordinates are well suited for identifying scaling solutions. For an interacting ground state of a harmonic trap that is then allowed to expand in a time dependent harmonic trap $V(x_\ii,t)\equiv m\omega^2_\ii(t)(x^\ii)^2/2$, we should expect to recover the scaling solution of Castin-Dum~\cite{castin_bose-einstein_1996}. We make a Thomas-Fermi ansatz for the particle density, which is time-independent in the scaling coordinates. The TF-ansatz initially has parabolic density
\begin{align} 
  |\phi(\rho,0)|^2&=\frac{\mu}{g N}\mathrm{max}\left(1-\frac{\rho_x^2}{R_x^2}-\frac{\rho_y^2}{R_y^2}-\frac{\rho_z^2}{R_z^2} ,0\right),
\end{align}
with radii $R_\jj^2=2\mu/m\omega^2_\jj(0)$, and is by definition time-independent in the scaling coordinates. For geometric mean $\bar R=(R_xR_yR_z)^{1/3}$, the TF-solution has 
\begin{align}  
N&=\mu\bar R^38\pi/(15g ),\\
a_\ii^2&= R_\ii^2/7,\\ \int d^3\rho\;|\phi|^4&=4\mu /(7g).
\end{align}
Using these properties, and applying the TF-approximation by neglecting the kinetic term in \eref{dsigdt}, we recover the Castin-Dum scaling equations~\cite{castin_bose-einstein_1996} in the form
\begin{align}
  \frac{d\sigma_\ii}{dt}=\frac{\omega^2_\ii(0)}{\lambda_\ii\bar\lambda^3}-\omega^2_\ii(t)\lambda_\ii.
\end{align}
We numerically integrate the scaling dynamics equations Eqs.~(\ref{dlamdt})-(\ref{dphidt}) for an initial condition in the TF-regime $\mu\gg \hbar\omega_\ii$, and prolate geometry $\omega_x=\omega_y\gg \omega_z$; in this case an analytical expression for the scalings is known for free expansion. To test this regime, we evolve a system with $(\omega_x,\omega_y,\omega_z)=(15,15,1)$, working again in units of $\omega_z$, $g_3 = 0.1\hbar\omega_z a_z^3$, $\mu = 25\hbar\omega_z$ well into the Thomas-Fermi regime. We use an initial domain $(L_x,L_y,L_z)=3(R_x,R_y,R_z)$ with $N_x=N_y=N_z=64$ points.

\begin{figure}[!t]
  \centering
\includegraphics[width=\columnwidth]{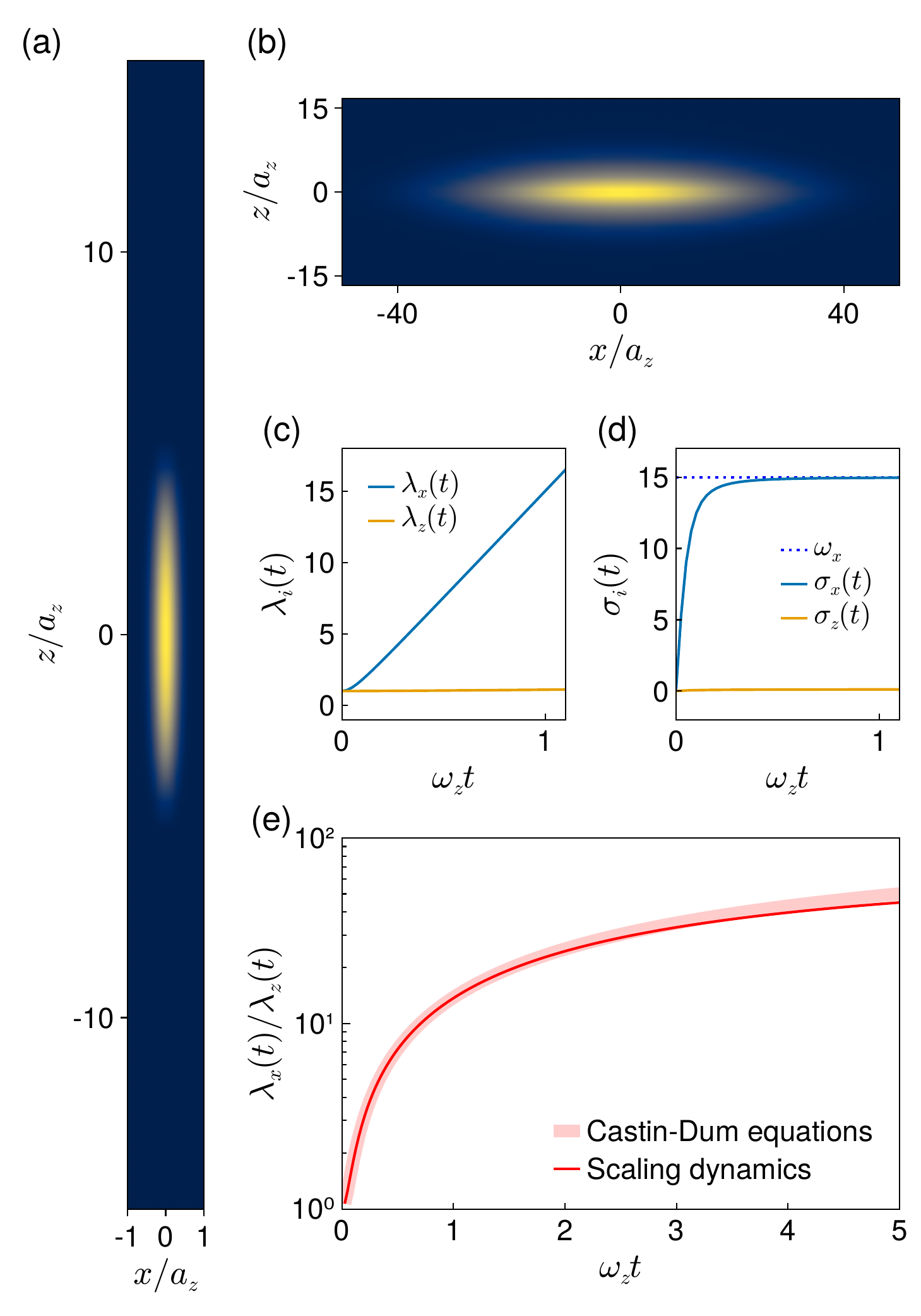}
\caption{Simulated expansion dynamics in the Thomas-Fermi regime on a grid of $64^3$ points. A GPE ground state in a harmonic trap with $(\omega_x,\omega_y,\omega_z)=(15,15,1)$ has initial density in the $x-z$ plane shown in (a), with large $\mu$. The density evolves to (b) after free expansion for $t=5/\omega_z\gg 1/\omega_x$. The scaling parameter $\lambda_x(t)$ rapidly becomes $\sim\omega_xt$, as seen in (c), expanding by a factor of $\sim 60$ by the final time. The $z$ evolution is almost frozen over the same time interval. (d) $\sigma_x(t)$ and $\sigma_y(t)$ rapidly approach their steady state $(\sigma_x,\sigma_y)\to(\omega_x,\omega_y)$. (e) The aspect ratio follows Castin-Dum scaling (Eqs. (20), (21) of Ref.~\cite{castin_bose-einstein_1996}) for high anisotropy at short times, departing slowly at longer times due to perturbative corrections to the anisotropic limit. 
}\label{fig2}
\end{figure}

The dynamics are presented in Fig.~\ref{fig2}. The expansion dynamics agrees closely with Castin-Dum analytic solution for a prolate system~\cite{castin_bose-einstein_1996}. In particular, the expansion rapidly approaches the scaling $\sigma_\ii\to \omega_\ii$, and the ratio $\lambda_z
(t)/\lambda_x(t)$ follows the scaling predicted analytically for highly elongated cigar traps.

\subsection{Strong antitrapping: position distribution}\label{sec:anti}
Consider a system that is subject to a time-dependent antitrapping potential~\cite{lewandowski_simplified_2003,coddington_experimental_2004}, of the form 
\begin{align}
  V(x_\ii,t)&\equiv-\frac{m}{2}\Omega^2_\ii(t)(x^\ii)^2.
\end{align}
For sufficiently large antitrapping potential energy the kinetic and interaction terms may be neglected from \eref{dsigdt}, and the scaling parameters evolve according to \begin{align} \frac{d\sigma_\ii(t)}{dt}&=\Omega_\ii(t)^2\lambda_\ii(t).
\end{align} 
The system state plays no role in the scaling as the external trap dominates the dynamics.
Remarkably, for such parabolic potentials the gauge potential in \eref{dphidt} is exactly cancelled~\footnote{Note that confining harmonic potentials will also cancel the gauge field, with strong confinement dynamics $\dot\sigma_\ii(t)=-\omega^2_\ii(t)\lambda_\ii(t)$, corresponding to a contraction scaling.}, leading to simplified scaling dynamics. 

For time-independent antitrapping in this regime, the scaling is exponential, 
\begin{align}\label{lmut}
  \lambda_\ii(t)=e^{\Omega_\ii t},
\end{align}
 with GPE dynamics 
\begin{align}
  i\hbar\frac{\partial \phi}{\partial t}&=\left(-\frac{\hbar^{2} \partial_{\ii}^{2}}{2 m \lambda^\ii(t)^2}+\frac{g_d N}{\bar{\lambda}(t)^\qq}|\phi|^{2}\right) \phi.\label{dphiadt}
\end{align}
For $d=3$ the interaction term vanishes rapidly, followed by the kinetic term.  For $d=2$  with symmetric antitrapping $\Omega_z=0, \Omega_x=\Omega_y= \Omega$, $\lambda_x(t)=\lambda_y(t)=\bar\lambda(t)$ and the kinetic and interaction terms have the identical scaling; 
as discussed in Appendix~\ref{app:b}, isotropic scaling can be absorbed into an effective time increment $ds(t)\equiv dt/\lambda(t)^2$. After physical time $t$ the GPE wavefunction evolves according to Eq.~(\ref{dphiadt}), with effective evolution time 
\begin{align}
  s(t)&=\int_0^t\frac{dt'}{\lambda(t')^2}=\int_0^t e^{-2\Omega t'}dt'=\frac{1}{2\Omega}(1-e^{-2\Omega t}).
\end{align}
Hence as $t\to\infty$, 
$s(t)\to(2\Omega)^{-1}$. Provided $\hbar\Omega\gg \mu$\;\footnote{Equivalent to the condition that the kinetic and interaction terms are negligible in \eref{dsigdt}.}, the GPE dynamics will be essentially frozen while the system size increases exponentially. Finally, we note that for $d=1$, $\bar\lambda(t)=\lambda_x(t)$, and the kinetic term vanishes rapidly, followed by the interaction term; expansion in a narrow tube is dominated by interactions before all dynamics are frozen.
\subsection{Aspect-ratio invariance: linear evolution}\label{sse}
The SDE can be used to identify interesting regimes of evolution for the scaling parameters $\lambda_\ii(t)$, as these are the size parameters of the condensate. 

We can seek a class of aspect-ratio invariant solutions with aspect ratios that are preserved under expansion. This is an approximate self-similar evolution that is directly accessible in experiments through observations of the cloud widths, it has also been interpreted as a signature of isotropic 3D quantum turbulence~\cite{henn_emergence_2009}. These states are distinct from self-similar dynamics introduced in STA in that the system shape parameters maintain their relative sizes, rather than the state preserving a particular functional dependence on scaling parameters.

For simplicity we will ignore initial changes in the scaling parameters ($\sigma_i(0)\equiv 0$). The dynamics can be analyzed simply in the lab frame coordinates. As shown in Appendix \ref{app:b}, the ideal gas evolves according to
\begin{align}\label{dsdti}
\frac{d\sigma_\ii}{dt}&=\frac{\tau_\ii-\sigma_\ii^2}{\lambda_\ii},
\end{align}
where the three independent stress tensor elements $\tau_\ii$ are  
 \begin{align}\label{tausimp2}
  \tau_\ii&=\frac{\hbar^2}{m^2a_\ii^2}\int d^\qq x\;|\partial_\ii\psi|^2=\frac{2K_\ii}{ma_\ii^2}\equiv \beta_i^2.
\end{align}
Here the kinetic energies $K_\ii$ are independent constants of the motion, and in terms of the momentum widths, $b_\ii\equiv \langle p_\ii^2\rangle^{1/2}$,
\begin{align} \label{beta}
  \beta_i &= \frac{b_\ii}{ma_\ii}
\end{align} 
are frequencies.

For aspect-ratio invariance \eref{dsdti} is required to be isotropic:
 \begin{align}
\frac{d}{dt}(\lambda_\ii\dot\lambda_\ii)=\tau_\ii\equiv \beta^2,
 \end{align}
with solution that preserves aspect ratios
\begin{align}\label{arats}
  \lambda_\ii(t)&=\sqrt{1+(\beta t)^2}.
\end{align}
The expansion timescale $1/\beta$ causes asymptotic rate  $\dot\lambda_\ii(t)\to\beta$ to be reached when $\beta t\gg 1$. The isotropy required for linear self-similar expansion, that $b_\ii/a_\ii$ are all equal,
\begin{figure}[!t]
  \centering
\includegraphics[width=\columnwidth]{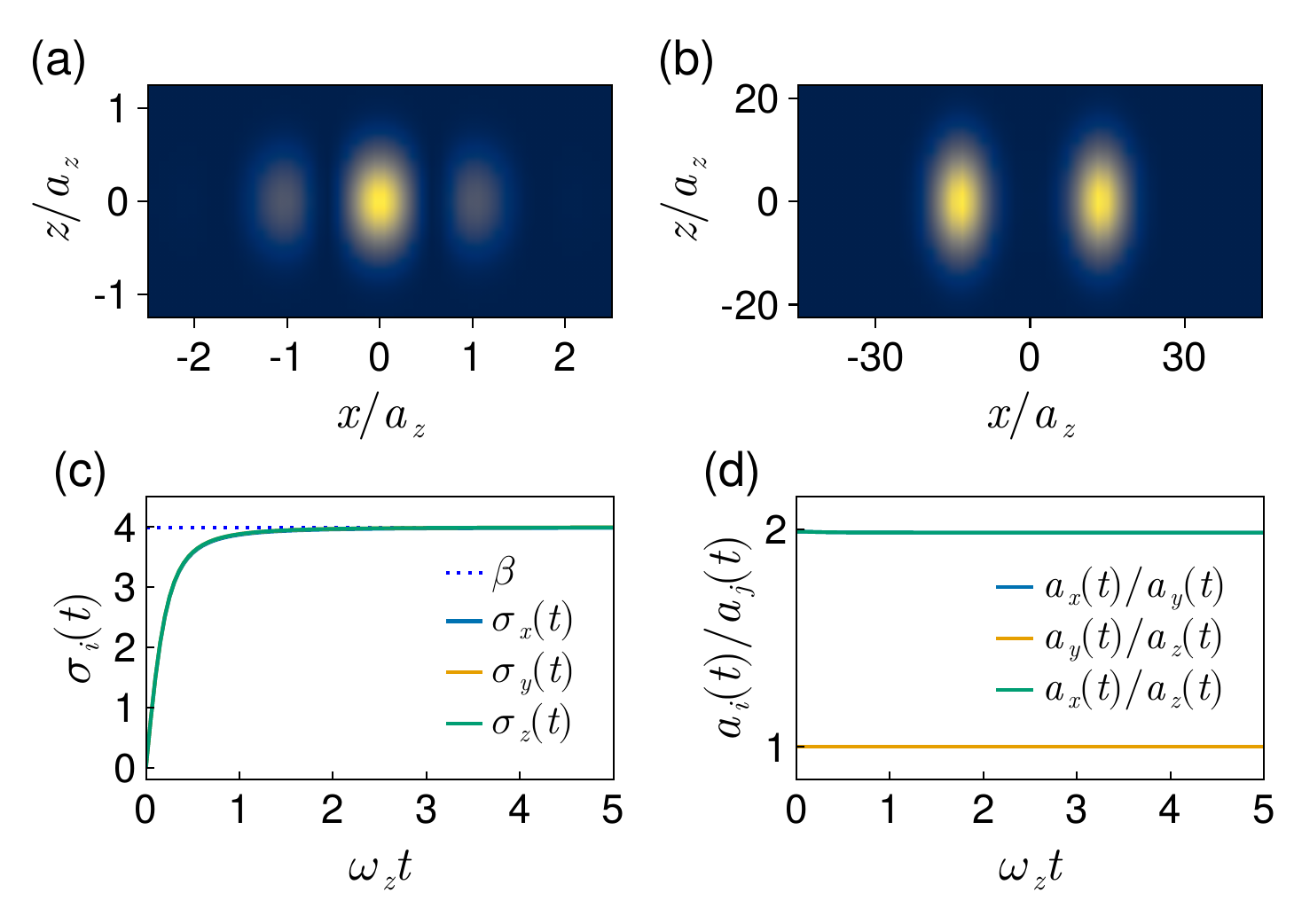}
\caption{Simulation of aspect-ratio invariant expansion from a superposition of counter-propagating trap ground states. The trap frequencies are $(\omega_x,\omega_y,\omega_z)=(1,4,4)$, and the superposition of translating ground states increases the momentum width in the $x$ direction to create isotropic $\beta=3.99$. (a) The initial position distribution is shown on the $x-z$ plane, where interference along $x$ is evident. (b) The final state is a scaled version of the initial state, consisting the two wavepackets located at $\pm k$ (see text). (c) The scaling parameters evolve isotropically, and the system maintains constant aspect ratio (d).
}
\label{fig3}
\end{figure} 
can be re-written as the set of conditions
\begin{align}\label{iso2}
  \frac{b_i}{b_j}=\frac{a_i}{a_j},\quad i\neq j.
\end{align}
Such a property is not exhibited by a harmonic trap ground state, for which the aspect ratios are inverses; a state satisfying \eref
{iso2} will thus involve a nontrivial superposition of trap eigenstates. 

As a simple test of linear aspect-invariant expansion, we evolve a system with $\omega_x:\omega_y:\omega_z=1:4:4$, consisting of the trap ground state in $y$ and $z$ directions. In the $x$ direction, the weaker trap means a wider position distribution, requiring a wider momentum distribution for isotropy of quadratic moments. We create a superposition of ground states in each direction, translating in the $\pm x$ direction. Again, in space and time units of $\sqrt{\hbar/m\omega_z}$, and $1/\omega_z$ respectively, for the $x$ part of the separable wavefuction, $\psi(x)$, we set $\psi(x)=\phi_0(x)(e^{iq x}+e^{-iq x})/\sqrt{2}$, having similar position width as the trap ground state $\phi_0(x)$, but momentum width increased by $\sim q$. For $q\gg \Delta k_0=1/\sqrt{2}$, the ground state momentum width, the width of the superposition approaches $\Delta k\simeq\sqrt{\Delta k_0^2+q^2}$. The momentum-space wavefunction consists of two ground-state wavepackets shifted to $\pm q$. Choosing $k=2.719$ gives $\Delta k=2.791$ and for our chosen trap frequencies, isotropic $\beta_i=3.99$. 

The results are shown in \fref{fig3}, where the initial state exhibits interference in the $x$ direction, and the final state is a scaled version of the initial momentum distribution, consisting of two separated wavepackets located at $\pm k$. 
\subsection{Aspect-invariance: nonlinear evolution}
Now that we have identified a linear aspect-invariant regime, how is this picture modified by contact interactions? For free expansion in 3D the scaling equation reads
\begin{align}\label{dsdt3d}
  \frac{d\sigma_\ii}{dt}&=\frac{1}{\lambda_\ii(t) m a_\ii^2}\int d^3\rho\;\left(\frac{g N}{2}\frac{|\phi|^4}{\bar\lambda(t)^3}+\frac{\hbar^2}{m\lambda_\ii(t)^2}|\partial_\ii\phi|^2 \right),
\end{align}
where both terms evolve in time, and transforming to scaling coordinates removes the $-\sigma_\ii^2$ term from \eref{dsdti}. These two changes, together with the interactions hamper analytic solution. However, we can use this formulation to gain insight into aspect-invariant dynamics. 

We proceed by assuming a non-trivial aspect-invariant solution exists. Starting from an anisotropic initial state ($a_i$ not all equal) the solution is required to evolve isotropically: $\lambda_i(t)\equiv \lambda(t)$. Such dynamics will occur provided there is an isotropic form of the equations of motion $d\sigma_i/dt\equiv d\sigma/dt$, or a consistent solution to the equation
\begin{align}\label{sexp}
  \frac{d\sigma }{dt}&=\frac{1}{\lambda(t)^3 }\int d^3\rho\;\left(\frac{g N}{2m a_\ii^2}\frac{|\phi|^4}{\lambda(t)}+\frac{\hbar^2}{m^2 a_\ii^2}|\partial_\ii\phi|^2 \right).
\end{align}
Since each term is positive definite,  a regime of self-consistent aspect-invariant dynamics exists only if  each term can be written in a form independent of $\ii$. As shown in the previous subsection the linear term can be isotropic provided $\beta_i\equiv \beta$, are independent of $\ii$. However, due to the anisotropic $a_\ii^2$ denominator, the otherwise isotropic interaction term cannot be cast into isotropic form. We thus arrive at a contradiction: for $s$-wave interactions an exact aspect-invariant solution of the GPE does not exist.

\section{Discussion and Outlook}\label{sec:outlook}
\subsection{Discussion}
We can note that the condition for \emph{linear} aspect-invariant expansion, \eref{iso2}, is in sharp contrast with a common interpretation of self-similar expansion in the context of nonlinear quantum turbulent BEC~\cite{henn_emergence_2009}. It was proposed that self-similar expansion is a consequence of an isotropic momentum distribution associated with well-developed quantum turbulence. However, momentum isotropy in a spatially isotropic system would violate \eref{iso2}, unless it is spherically symmetric in both position and momentum space --- an uninteresting special case. We conclude that a state with momentum isotropy and position anisotropy cannot undergo exact nonlinear  aspect-invariant expansion. 
Approximate aspect-invariant dynamics may occur if isotropic $\beta_i$ condition holds and the interaction term decays faster than the kinetic term, possible in the regime of rapid initial expansion due to the $\bar\lambda(t)^{-3}$ interaction-term scaling in \eeref{dsdt3d}.   A system could enter a regime of approximate \emph{linear} aspect-ratio invariance if the interaction energy is much smaller than the kinetic energy, provided that the average momentum per unit length, as defined in \eref{beta}, is isotropic. 
States with relatively high initial kinetic energy could more easily enter this regime, which may offer a partial explanation for the observations in Ref.~\cite{henn_emergence_2009} for high-energy turbulent states.

Another possible explanation of nonlinear aspect-invariance reported in Ref.~\cite{henn_emergence_2009} was put forward by Caracanhas \emph{et al.}~\cite{caracanhas_self-similar_2012,caracanhas_self-similar_2013-1}. A semiclassical rotational velocity field was introduced within a variational ansatz, and shown to cause approximate aspect-invariant expansion under specific conditions of alignment with tht trap anisotropy. Such a rotational field would be associated with long-range velocity coherence in the superfluid. The relevance of this mechanism for aspect-invariant expansion and quantum turbulence remains an open problem.

Returning to the equations of motion, in principle \eref{dphidt} alone could be used to simulate the entire expansion dynamics without the need for extremely large numerical grids. However, without \eref{dsigdt} there is a significant limitation as the scalings are not known \emph{apriori}. A number of works have approached this problem by various analytic approximations for $\lambda_\ii(t)$, including the Castin-Dum scaling solution~\cite{castin_bose-einstein_1996}, the hydrodynamic self-similarity ansatz~\cite{guery-odelin_mean-field_2002,modugno_effective_2018,viedma_effective_2020}, or a linear approximation for the scaling parameter time dependence~\cite{deuar_tractable_2016}. The exact equation of motion for $\lambda_\ii(t)$ found here is able to accommodate the evolution of arbitrary initial states and external potentials. 

\subsection{Outlook}
In this work we have recast the dynamics of a cold BEC described by the GPE in scaling coordinates determined by the size parameters of the system. The scaling GPE  self-consistently adapts as the system size evolves, driven by diagonal elements of the quantum fluid stress tensor. We have verified our formulation by evolving three-dimensional systems in the ideal gas and Thomas-Fermi regimes, observing close agreement with known analytical results. We identified a linear regime of aspect-ratio preserving expansion for spatially anisotropic states. The aspect-invariant dynamics was shown to be a consequence of the system having identical aspect ratios in position and momentum space, equivalent to isotropic average momentum per unit length. Numerical simulation of a superposition of counter-propagating ground state wavepackets verified this aspect-invariant expansion condition for the non-interacting gas. Analysis of the nonlinear scaling dynamics shows that there are no solutions with exact aspect-invariance under free expansion. 

An interesting future direction would be to further investigate conditions for aspect-invariant expansion and its precise connection to different states of 3D quantum turbulence~\cite{henn_emergence_2009}. The expansion problem is also closely linked to construction of shortcuts to adiabaticity~\cite{del_campo_shortcuts_2013,guery-odelin_shortcuts_2019} involving matter wave manipulation over large spatial and temporal scales. The scaling GPE enables large-scale numerical modelling of shortcuts by simulating the full matter wave evolution. As shortcuts are sometimes easily constructed for a subset of system parameters, the scaling GPE could be used to investigate uncontrolled degrees of freedom excited during a shortcut, such as the parasitic excitations reported in \cite{schaff_shortcut_2011}.

Finally we note that as our treatment only relies on the fluid continuity equation and stress tensor, fairly generic properties of cold quantum gases, a fruitful direction would be to extend the scaling GPE approach to more exotic interactions, to systems of Fermions or Bose-Fermi mixtures, and to settings beyond the scope of mean field theory. 

\acknowledgments
We thank Xiaoquan Yu, Amita Deb, and Danny Baillie for stimulating discussions. AB acknowledges support from the Marsden Fund (Grant No. UOO1726) and the Dodd-Walls Centre for Photonic and Quantum Technologies.

\appendix
\section{Scaling GPE}\label{app:a}
In the scaling coordinates $\partial_\ii=\partial/\partial \rho_\ii$, we can find the 
transformed derivative 
\begin{align} 
  D_\ii&\equiv e^{-im \rho_\jj \rho^\jj\lambda^\jj\dot\lambda_\jj/2\hbar}\partial_\ii e^{im  \rho_\sigma \rho^\sigma\lambda^\sigma\dot\lambda_\sigma/2\hbar},
\end{align} 
and Laplacian $D^\ii D_\ii$ in the form
\begin{align}\label{dmu}
D_\ii&=\partial_\ii+i\frac{m}{\hbar}\lambda_\ii\dot\lambda_\ii\rho_\ii,\\\label{ddmu}
D^\ii D_\ii&=\partial^\ii\del_\ii+i\frac{m}{\hbar}\dot\lambda_\ii\lambda^\ii(2\rho^\ii\partial_\ii+1)-\frac{m^2}{\hbar^2}(\dot\lambda_\ii\lambda^\ii)^2\rho_\ii\rho^\ii.
\end{align}
Using the definition \eref{psiscale}, in lab frame coordinates the GPE reads
\begin{align}
i\hbar\frac{\partial \psi}{\partial t}&=\Bigg(i\hbar\frac{\partial \phi}{\partial t}+i\hbar\bar\lambda^{\qq/2}\left(\frac{d}{dt}\frac{1}{\bar\lambda^{\qq/2}}\right)\phi -\frac{m}{2}x^\ii  x_\ii\left(\frac{d}{dt}\frac{\dot\lambda_\ii}{\lambda^\ii}\right)\phi \notag\\
&-i\hbar\frac{\dot\lambda_\ii}{(\lambda^\ii)^2}x^\ii\partial_\ii\phi\Bigg)\frac{e^{im \dot\lambda_\jj x_\jj x^\jj/2\hbar\lambda^\jj}}{\bar\lambda^{\qq/2}}\notag\\
&=\left(-\frac{\hbar^2\partial_\ii^2}{2m}+V(x^\ii,t)+g|\psi|^2\right)\psi.
\end{align}
In the scaling coordinates this can be written as
\begin{align}\label{gpe2}
i\hbar\frac{\partial \phi}{\partial t}&=\Bigg[V(\rho^\ii\lambda^\ii,t)+\frac{g}{\bar\lambda^\qq}|\phi|^2+\frac{m}{2}\lambda_\ii\lambda^\ii\rho^\ii  \rho_\ii\frac{d}{dt}\left(\frac{\dot\lambda_\ii}{\lambda^\ii}\right)\notag\\
&-i\hbar\bar\lambda^{\qq/2}\frac{d}{dt}\frac{1}{\bar\lambda^{\qq/2}}+i\hbar\frac{\dot\lambda_\ii}{\lambda^\ii}\rho^\ii\partial_\ii\Bigg]\phi \notag\\
&-\frac{\hbar^2}{2m\lambda^\ii(t)^2}(e^{-im \rho_\jj \rho^\jj\lambda^\jj\dot\lambda_\jj/2\hbar}\partial_\ii^2e^{im  \rho_\sigma \rho^\sigma\lambda^\sigma\dot\lambda_\sigma/2\hbar})\phi.
\end{align}
Using the identities
\begin{align}
\frac{d}{dt} \frac{\dot\lambda_\ii}{\lambda^\ii} &=\frac{\ddot\lambda_\ii}{\lambda^\ii}-\left(\frac{\dot\lambda_\ii}{\lambda^\ii}\right)^2,\\
\bar\lambda^{\qq/2}\left(\frac{d}{dt}\frac{1}{\bar\lambda^{\qq/2}}\right)&=-\frac{1}{2}\frac{\dot\lambda_\ii}{\lambda^\ii}.
\end{align}
and the Laplacian, (\ref{ddmu}), after significant cancellation of terms, we find $\phi(\rho_\ii,t)$ evolves according to \eref{dphidt}.
\section{Free expansion of the ideal gas}\label{app:b}
In the simplest non-interacting scenario, the evolution is governed by the one-body Hamiltonian
\begin{align}
\hat H_0&=\frac{\hat p^2}{2m},
\end{align}
for momentum operator $\hat p=(\hat p_x,\hat p_y,\hat p_z)$.

The wavefunction $\psi(\rr,0)$, with momentum-space representation
\begin{align}\label{kspacewf}
\tilde\psi(\bpp,0)&\equiv\frac{1}{(2\pi\hbar)^d}\int d^d r\;e^{-i\bpp\cdot\rr/\hbar}\psi(\rr,0).
\end{align}
evolves to 
\begin{align}
\tilde\psi(\bpp,t)&=\exp{\left(-i\frac{ p^2t}{2m\hbar}\right)}\tilde\psi(\bpp,0).
\end{align}
The density
\begin{align}
|\psi(\rr,t)|^2&=\int d^d p\int d^d p'\frac{e^{i(p^2-p'^2)t/(2m\hbar) +i(\bpp'-\bpp)\cdot\rr/\hbar}}{(2\pi\hbar)^d}\notag\\\label{dens1}
&\times\tilde\psi^*(\bpp,0)\tilde\psi(\bpp',0).
\end{align}
can be written more usefully with the change of variables $\bPP = (\bpp+\bpp')/2 $, $\bqq=\bpp'-\bpp$ as
\begin{align}\label{desn2}
|\psi(\rr,t)|^2&=\int d^dP\int d^dq\frac{e^{i\bqq\cdot(\rr-\bPP t/m)/\hbar}}{(2\pi\hbar)^d}\notag\\
&\times\tilde\psi^*(\bPP- \bqq/2,0)\tilde\psi(\bPP+ \bqq/2,0).
\end{align}
For $t>0$ we can use the scaled variable $\bxx = \bqq t/m$, giving
\begin{align}\label{desn3}
|\psi(\rr,t)|^2&=\left(\frac{m}{ t}\right)^d\int d^dP\int d^dx\frac{e^{i\bxx\cdot(\bPP-m\rr/ t)/\hbar}}{(2\pi \hbar)^d}\notag\\
&\times\tilde\psi^*\left(\bPP+\frac{m\bxx}{2 t},0\right)\tilde\psi\left(\bPP-\frac{m\bxx}{2 t},0\right).
\end{align}

For long times, a Taylor series in $m\bxx/2 t$ can be truncated at lowest order, and the $x$ integral
\begin{align}\label{desn5}
  \int d^dx\; e^{i\bxx\cdot(\bPP-m\rr/ t)/\hbar}&=(2\pi\hbar)^d\delta^{(d)}(\bPP-m\rr/t)
\end{align}
gives the long-time limit of the particle density
\begin{align}\label{freexp}
\lim_{t\to\infty}|\psi(\rr,t)|^2&=\left(\frac{m}{t}\right)^d\Big|\tilde\psi\left(\frac{m\rr}{t},0\right)\Big|^2,
\end{align}
in terms of the initial momentum distribution evaluated at the rescaled momentum $m\rr/t$. The long-time limit provides a useful test of our reformulation of the Gross-Pitaevskii equation.

We can solve the free expansion problem exactly for the case of the ideal Bose gas, to gain further insight into the scaling GPE. Without interactions and external potential, we have 
\begin{align}
  \frac{d\sigma_\ii}{dt}&=\frac{1}{\lambda_\ii(t)^3ma_\ii^2}\int d^\qq\rho\frac{\hbar^2}{m}|\partial_\ii\phi|^2,\\\label{idealse}
  i\hbar\frac{\partial \phi}{\partial t}&=\left(-\frac{\hbar^{2} \partial_{\ii}^{2}}{2 m \lambda^\ii(t)^2}+\frac{m}{2}  \dot{\sigma}_{\ii}(t) \lambda^{\ii}(t)\rho_\ii\rho^\ii\right) \phi,
\end{align}
where there is no obvious analytical solution. However, in lab frame coordinates, the $\sigma_\ii(t)$ equation is 
\begin{align}\label{sigddfree}
  \frac{d\sigma_\ii}{dt}&=\frac{\tau_\ii-\sigma_\ii^2}{\lambda_\ii},
\end{align}
where 
\begin{align}
  \tau_\ii&=\frac{\hbar^2}{m^2a_\ii^2}\int d^\qq x\;|\partial_\ii\psi|^2=\frac{\langle p_\ii^2\rangle}{m^2a_\ii^2}
\end{align} 
is a constant of the motion. We can recast \eref{sigddfree} as 
\begin{align}\label{int1}
  \lambda_\ii\ddot\lambda_\ii+\dot\lambda_\ii^2\equiv \frac{d(\lambda_\ii\dot\lambda_\ii)}{dt}&=\tau_\ii,
\end{align}
and integrate with initial condition $\dot\lambda_\ii(0)=\sigma_\ii(0)$ to give
\begin{align}\label{int2}
\lambda_\ii\dot\lambda_\ii&=\sigma_\ii(0)+\tau_\ii t,
\end{align}
and integrate again with initial condition $\lambda_i(0)=1$ gives the general solution 
\begin{align}\label{int3}
  \lambda_\ii(t)&=[1+2\sigma_\ii(0)t+\tau_\ii t^2]^{1/2}.
\end{align} 
We can now simplify the scaling GPE for free expansion by using equations of motion \eeref{int2},\eeref{int3} to find
\begin{align}
  \dot\sigma_\ii\lambda_\ii&=\frac{\tau_\ii-\sigma_\ii(0)^2}{\lambda_\ii^2}.
\end{align}
The scaling Schr\"{o}dinger equation \eeref{idealse} hence reduces to 
\begin{align}\label{idealser} 
  i\hbar\frac{\partial \phi}{\partial t}&=\frac{1}{\lambda_\ii(t)^2}\left(-\frac{\hbar^{2} \partial_{\ii}^{2}}{2 m}+\frac{m}{2} \alpha_\ii^2\rho_\ii\rho^\ii\right) \phi,
\end{align}
where $\alpha_\ii\equiv \sqrt{\tau_\ii-\sigma_\ii(0)^2}$ determines whether the effective harmonic trap is attractive, repulsive, or vanishing.  
There are thus three cases to consider depending on the size of the initial momentum widths and accelerations: momentum dominated regime, acceleration dominated regime, and critical regime. 

We can characterize the dynamics of \eref{idealser} by noting that direction $\ii$ evolves with an effective time increment $
  ds_\ii(t) \equiv dt/\lambda_\ii(t)^2$. Physical time evolution $t$ is thus equivalent to axis-dependent effective time
\begin{align}
  s_\ii(t)&=\int_0^t\frac{du}{\lambda_\ii(u)^2}=\int_0^t\frac{du}{1+2\sigma_\ii(0)u+\tau_\ii u^2}.
\end{align}
We now consider three cases of the dynamics, according to the discriminant of the denominator, $\alpha_\ii$.
\begin{enumerate}
  \item \emph{Momentum dominated regime:} $\alpha_\ii^2>0$ and real valued $\alpha_\ii= \sqrt{\tau_\ii-\sigma_\ii(0)^2}$, with solution
  \begin{align}
    s_i(t)=\frac{1}{\alpha_\ii}\arctan{\left(\frac{\alpha_\ii t}{1+\sigma_\ii(0) t}\right)}.
  \end{align}
  In this most common case $\alpha_\ii$ is an effective trapping frequency. We can check that the long time limit is physical by noting that after long free expansion times, $t\gg 1/\alpha_\ii$, the total effective evolution time in the harmonic potential is $s_\ii(t)\to\pi/(2\alpha_\ii)$, one quarter period at frequency $\alpha_\ii$, independent of the initial stress $\tau_\ii$. Since one quarter period transforms a position distribution into its momentum distribution in a harmonic trap, in this regime we recover the well-known long-time limit of free expansion, as discussed in Section \ref{app:b}. 
  \item \emph{Acceleration dominated regime:} $\alpha_\ii^2<0$ and real valued $\tilde\alpha_\ii \equiv\sqrt{\sigma_\ii(0)^2-\tau_\ii}$, with solution
  \begin{align}
    s_i(t)=\frac{1}{\tilde\alpha_\ii}\operatorname{arctanh}{\left(\frac{\tilde\alpha_\ii t}{1+\sigma_\ii(0) t}\right)},
  \end{align}
  corresponding to antitrapping with effective trap frequency $\tilde\alpha_\ii$. This regime has a very different long-time limit. When $t\gg 1/\tilde\alpha_\ii $, the effective time approaches $s_\ii(t)\to \infty$ and the evolution with large initial acceleration is simply the long time limit of antitrapped expansion. 
  \item \emph{Balanced regime:} $\alpha_\ii=0$, and $\sigma_\ii(0)=\sqrt{\tau_\ii}$, with solution
  \begin{align}
    s_i(t)&=\frac{t}{1+\sqrt{\tau_\ii}t}.
  \end{align}
 The balance of initial acceleration and momentum width causes the effective trap in \eref{idealser} to vanish and the scaling dynamics proceeds according to the free-space Schr\"{o}dinger equation. In the long-time limit, $t\gg 1/\sqrt{\tau_\ii}$, the effective evolution time approaches $s_\ii(t)\to 1/\sqrt{\tau_\ii}$.
\end{enumerate}
We have verified that the scaling dynamics equations reproduce the well-known result for free expansion in the momentum-dominated regime: long time-of-flight maps the momentum distribution into the position distribution.


\providecommand{\noopsort}[1]{}

\end{document}